\documentclass[doublecol]{epl2} % for 2 columns style with line numbers
\title{Polyakov--Nambu--Jona-Lasinio model in finite volumes}

\author{Abhijit Bhattacharyya\inst{1}\footnote{abhattacharyyacu@gmail.com}
\and Sanjay K. Ghosh\inst{2,3}\footnote{sanjay@jcbose.ac.in} 
\and Rajarshi Ray\inst{2,3}\footnote{rajarshi@jcbose.ac.in} 
\and Kinkar Saha\inst{2,3}\footnote{saha.k.09@gmail.com} 
\and Sudipa Upadhaya\inst{2,3}\footnote{sudipa.09@gmail.com} }
\shortauthor{A. Bhattacharyya \etal}

\institute{                    
  \inst{1} Department of Physics, University of Calcutta,
             92, A.P.C Road, Kolkata-700009, INDIA\\
  \inst{2} Department of Physics, Bose Institute, 
	  93/1, A. P. C Road, Kolkata - 700009, INDIA,Center for Astroparticle Physics \&
Space Science, Block-EN, Sector-V, Salt Lake, Kolkata-700091, INDIA\\
 \inst{3} Center for Astroparticle Physics \&
Space Science, Block-EN, Sector-V, Salt Lake, Kolkata-700091, INDIA 
}

\abstract{
We discuss the 2+1 flavor Polyakov loop enhanced Nambu--Jona-Lasinio model 
in a finite volume. The main objective is to check the volume scaling
of thermodynamic observables for various temperatures and chemical
potentials. We observe the possible violation of the scaling with
system size in a considerable window along the whole transition region
in the $T-\mu_q$ plane.
}

\begin{document}

\maketitle

%***********************************************************************
\section{Introduction}

The hot and/or dense matter created in ultra-relativistic heavy ion
collisions is supposed to possess a rich phase structure. In the
intermediate regime of temperature and baryon chemical potential in the
range of few hundred MeV, the defining characters of the phases
are the color confinement and chiral properties. While for low baryon
densities the matter has a smooth crossover from color confined
chiral symmetry broken phase to color deconfined chiral symmetry restored
phase, at high enough densities, this transition may be of first order.
A critical end point seems to naturally occur in such a situation.
Establishing this scenario forms an integral part of exploration in the
international collaborative experiments at CERN and BNL and the upcoming
experiments at GSI.

The matter formed in heavy-ion collision experiments has a finite
volume, which depends on the size of the colliding nuclei as well as the
center of mass energy ($\sqrt s$) and the centrality of collisions.
Measurement of this system size is quite non-trivial. There are several
estimates of the system size at freeze-out for different $\sqrt s$ and
different centralities from the measurement of HBT radii~\cite{ceres},
which indicate that the freeze out volume increases as the $\sqrt s$
increases. The freeze out volume was found to be in the range of
$2000~fm^3$ to $3000~fm^3$.
On the other hand in the URQMD model~\cite{urqmd} the volume of
homogeneity has been calculated and compared with the experimentally
available results in Ref.~\cite{graf}. The homogeneity
volume~\cite{Sinyukov,Heinz} is given by the product of the HBT radii
$R_{out}$, $R_{side}$ and $R_{long}$, signifying the fraction of the
fireball from which particles in a particular momentum window is
emitted. The volume of homogeneity was obtained in the transverse
momentum window of 300-400 $MeV$ at different centralities and varying
$\sqrt{s}$ from 62.4 $GeV$ to 2760 $GeV$. These UrQMD simulations show
that the homogeneity volume has a power law scaling with multiplicity
and varies in the range of 50 to 250 $fm^3$. The ALICE collaboration has
also estimated the freeze-out volume to be in this range for different
$\sqrt s$ and different colliding nuclei~\cite{alice-fs}.
Given that these are the freeze-out volumes, the initial volumes are
expected to be much smaller~\cite{initi-fs1,initi-fs2}.

Theoretical developments to understand the effects of finite volume on
the strongly interacting matter is represented in
Refs.~\cite{balian,jaffe,candelas,elze,lusher1,gasser1,madsen,spieles,
fslat1,kiriyama,fischer1,shao,yasui,fslat2,fischer2,braun1,abreu1,
braun2,khanna,ebert,keliu,plsm,npqcd}.  General overview of the
developments in this direction is reviewed in some of our recent
work~\cite{Bhattacharyya1,Bhattacharyya2,Subhasis}.  In case of high
energy heavy-ion collisions it was shown in Ref.~\cite{fraga05} that the
finite size effects give rise to important consequences in the later
stage of evolution of the hadronic bubbles.  In Ref.~\cite{fraga11a}, the
authors have discussed the importance of finite size scaling in order to
properly locate the critical end point (CEP) on the phase diagram and
emphasized on its use for the experimental data analysis
program~\cite{fraga11b}.  In Ref.~\cite{Bhattacharyya1} some of us have
studied the thermodynamic properties of strongly interacting matter in a
finite volume using Polyakov-Nambu-Jona--Lasinio (PNJL) model for 2 and
2+1 flavors. It was found that the cross-over temperature at zero baryon
density decreases with decrease in volume. Furthermore the critical end
point at finite temperature $T$ and chemical potential $\mu$ goes
towards higher $\mu$ and lower $T$ domain as system size is reduced.
For a system confined to a lateral size $R=2fm$, it was found that the
CEP vanishes and the whole phase boundary becomes a cross-over line.  We
also discussed the possibility of chiral symmetry restoration in a color
confined state for a system with finite size.

Fluctuations of conserved charges are sensitive indicators of the
transition from hadronic matter to partonic state. Also the existence of
the CEP may be signalled by the diverging behavior of fluctuations.
From the chiral susceptibilities in the 2-flavor quark-meson model using
a renormalisation group approach~\cite{tripolt} the CEP was found to
have lower temperature and higher chemical potentials with decreasing
volume, similar to our results in the PNJL model~\cite{Bhattacharyya1}.
Similar agreement was reported from baryon number susceptibility ratios
via Dyson-Schwinger approach using 2-flavor system~\cite{npqcd}.  On the
lattice, for a pure gluon theory, the Polyakov loop susceptibility has
been calculated for a finite volume~\cite{berg} showing that the
transition temperature may increase with decreasing system size. In a
lattice version of the 2-flavor PNJL model~\cite{cristoforetti} the
diagonal quark number susceptibility was found to decrease with
increasing volume. Our recent studies in the mean-field approximation of
2-flavor PNJL model~\cite{Bhattacharyya2} as well as in hadron resonance
gas model~\cite{Subhasis}, give similar results. Additionally we found
that the volume scaling of susceptibility ratios is violated near the
cross-over region in the PNJL model and near the freeze-out surface in
the hadronic model. It is therefore important to explore the situation
in the 2+1 flavor PNJL model, which is reported in the present work.

We first give a brief description of the PNJL model followed by the
results for the fluctuations and correlations.

\section{Model}

The NJL model~\cite{YNambu,Kunihiro1,Vogl1,Klevansky1,Hatsuda1,Buballa1}
gives a satisfactory description of strongly interacting matter at zero
temperature and chemical potentials. However this is a model with global
color conservation and therefore no confinement of color charges appear
at non-zero temperatures and densities. A suitable modification is done
by introducing a background field that mimics the behavior of the
Polyakov loop to bring in the effects of confinement~\cite{Meisinger,
Fukushima1,Ratti1,Megias1,Ghosh1,SMukherjee1,Ghosh2,Tsai1,Megias2}.
Considerable progress has been made to have an understanding of strongly
interacting matter using the PNJL model (see e.g.~\cite{Abuki1,Abuki2,
Boomsma1,Bhat4,Sasaki1,Bhat5,Bhat7,KSaha1,Marty1,Dutra1,Ghosh3,
Xin1,Islam1,Ghosh4,SUpadhaya1,SUpadhaya2,Islam2,Bhattacharyya1,
Bhattacharyya2,SuKi}. Here we have considered 2+1 flavor PNJL model
including upto six-quark interactions~\cite{Bhattacharyya1,Saumen}. To
include the effect of finite volumes, a non-zero lower momentum cutoff
$p_{min}=\pi/R=\lambda$ is introduced. Here $R$ is the lateral size of a
cubic volume $V=R^3$. This is a simple approach as compared to more
sophisticated approaches needed to compute the proper density of states
as discussed for various systems in Ref.~\cite{balian,jaffe,candelas,
elze,lusher1,gasser1,madsen}. In principle there should be an infinite
sum over discrete momentum values. However for simplification, we
consider integration over continuous values of momentum. Alternatively, 
surface and curvature effects should have been included in the
continuous momentum variables which are beyond the scope of the present
work. The essential physical effects are expected to remain similar.
Most likely we are going to make an underestimation of finite size
effects in the quantities we evaluate. We have also not considered any
modification of the model parameters with the expectation that much of
those effects would show up in the mean values of the quantum fields.
This is exactly in the line of neglecting the temperature and chemical
potential dependence of model parameters in most of the related
literature. Within this approximation, the thermodynamic potential is
now given by~\cite{Bhattacharyya1,Saumen},

%%%%%%%%%%%%%%%%%%%%%%%%%%%%%%%%%%%%%%%%%%%%%%%%%%%%%%%%%%%%%%%%%%%%%%%%
\begin{eqnarray}
& \Omega =
  \mathcal{U'}(\Phi,\bar{\Phi},T)
 +2g_S\sum_{f=u,d,s}\sigma_f^2-\frac{g_D}{2}\sigma_u\sigma_d\sigma_s
&\nonumber \\
& +\sum_{f=u,d,s}\left[ 6\int_\lambda^\Lambda\frac{d^3p}
  {(2\pi)^2}E_p^f
       - 2 T\int_\lambda^\infty\frac{d^3p}{(2\pi)^3}
\right.
&\nonumber \\
& \left\{ln\left[1+3\left(\Phi+\bar{\Phi}
            e^{\frac{-(E_p^f-\mu_f)}{T}}\right)
            e^{\frac{-(E_p^f-\mu_f)}{T}}            
           +  e^{\frac{-3(E_p^f-\mu_f)}{T}} \right] \right.
&\nonumber \\
&  \left. \left.ln\left[1+3\left(\Phi+\bar{\Phi}
            e^{\frac{-(E_p^f+\mu_f)}{T}}\right)
            e^{\frac{-(E_p^f+\mu_f)}{T}} 
           + e^{\frac{-3(E_p^f+\mu_f)}{T}}\right] \right\} \right].
&
\end{eqnarray}
%%%%%%%%%%%%%%%%%%%%%%%%%%%%%%%%%%%%%%%%%%%%%%%%%%%%%%%%%%%%%%%%%%%%%%%%
Here $\Phi$ and $\bar{\Phi}$ are the Polyakov loop field and its
conjugate respectively. The condensate fields of different flavors
($f=u,d,s$) are given by $\sigma_f=\langle\bar{\psi_f}\psi_f\rangle=
-\frac{3M_f}{\pi^2}\int_\lambda^\Lambda \frac{p^2}{E_p^f}dp$, where
$E_p^f= \sqrt{p^2+M_f^2}$ is the corresponding quasiparticle energy with
quasiparticle mass $M_f =m_f-g_S \sigma_f+g_D \sigma_g\sigma_h$. The NJL
parameters are the 4-quark coupling $g_S$, the 6-quark coupling $g_D$
and the ultraviolet cut-off $\Lambda$. The Polyakov loop potential is
given by,
%%%%%%%%%%%%%%%%%%%%%%%%%%%%%%%%%%%%%%%%%%%%%%%%%%%%%%%%%%%%%%%%%%%%%%%%
\begin{eqnarray}
 \frac{\mathcal{U'}(\Phi,\bar{\Phi},T)}{T^4}=
-\frac{b_2(T)}{2}\bar{\Phi}\Phi-\frac{b_3}{6}(\Phi^3+\bar{\Phi}^3)
                 +\frac {b_4}{  4}{(\bar\Phi \Phi)}^2 \nonumber \\
-\kappa \ ln[1-6\Phi\bar{\Phi}+
4(\Phi^3+\bar{\Phi}^3)-3(\Phi\bar{\Phi})^2]
\label{Ppotential}
\end{eqnarray}
%%%%%%%%%%%%%%%%%%%%%%%%%%%%%%%%%%%%%%%%%%%%%%%%%%%%%%%%%%%%%%%%%%%%%%%%
where, $b_2(T)=a_0+a_1{\left(\frac{T_0}{T}\right)}+a_2
\left(\frac{T_0}{T} \right)^2+a_3\left(\frac{T_0}{T}\right)^3$. Here
$a_0$, $a_1$, $a_2$, $T_0$, $b_3$ and $b_4$ are constants which may be
obtained by fitting the temperature dependence of the Polyakov loop and
pressure in pure gauge theory in the lattice framework~\cite{Ratti1}.
The thermodynamic potential $\Omega$ is first extremized with respect to
the $\sigma$ and $\Phi$ fields to estimate the mean field values at
desired temperature and chemical potential. The values of the mean
fields are plugged back into $\Omega(T,\mu)$ to obtain the mean
thermodynamic potential.

The model parameters used here are ~\cite{Bhattacharyya1}, $a_0=6.75$,
$a_1=-1.95$, $a_2=2.625$, $a_3=-7.44$, $b_3=0.75$, $b_4=7.5$, $T_0=190
\, {\rm MeV}$, $\kappa = 0.13$, $m_u=m_d=5.5 \, {\rm MeV}$, $m_s=134.76
\, {\rm MeV}$, $\Lambda = 631 \, {\rm MeV}$, $g_s \Lambda^2 = 3.67$ 
and $g_D \Lambda^5= 9.33 $. The cross-over
temperatures for the different system sizes studied here is given in
table~\ref{tab.tc}.
%%####################################################################%%
\begin{table}[!htb]
\centering
\begin{tabular}{|c|c|c|c|c|}
\hline
 R(fm)  & 2 & 3 & 4 & $\infty$ \\
\hline
$T_c$(MeV) & 160 & 174 & 178 & 181     \\
\hline
\end{tabular}
\caption{Cross-over temperatures corresponding to  various system sizes
as extracted in 2+1-flavored PNJL model at vanishing chemical potential.}
\label{tab.tc}
\end{table}
%%####################################################################%%

\section{Results}

%%%%%%%%%%%%%%%%%%%%%%%%%%%%%%%%%%%%%%%%%%%%%%%%%%%%%%%%%%%%%%%%%%%%%%%%
\begin{figure*}[!htb]
\begin{center}
\includegraphics[height=4.0cm,width=4.3cm]{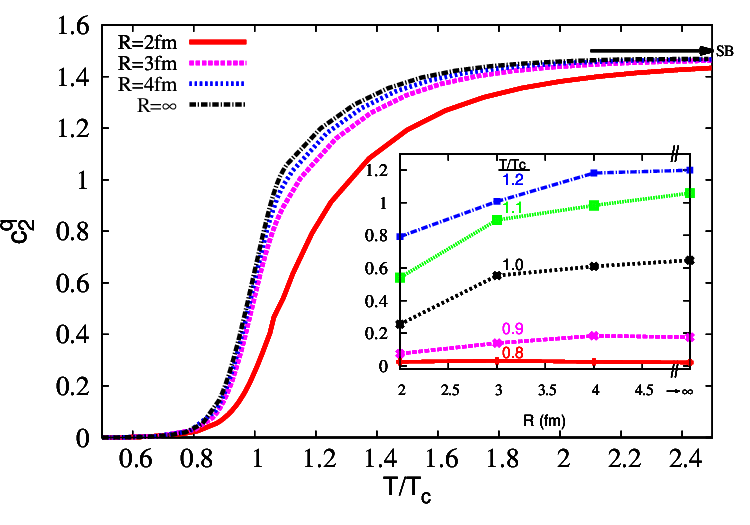}
\includegraphics[scale=0.45]{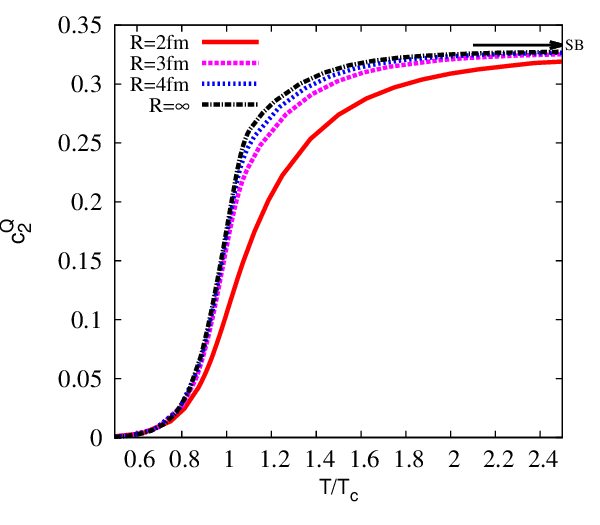}
\includegraphics[scale=0.45]{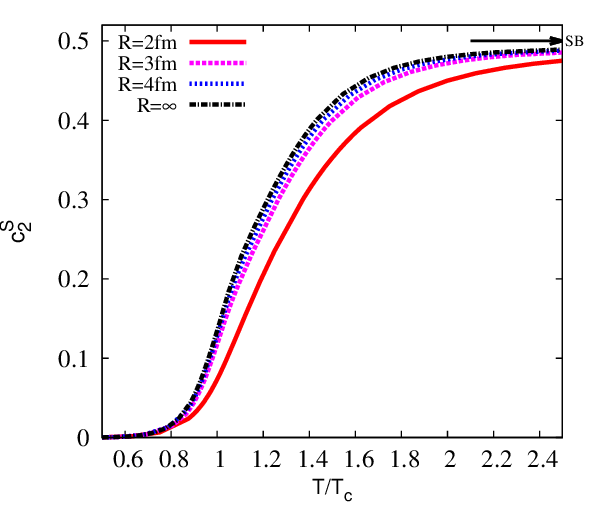}

\includegraphics[scale=0.45]{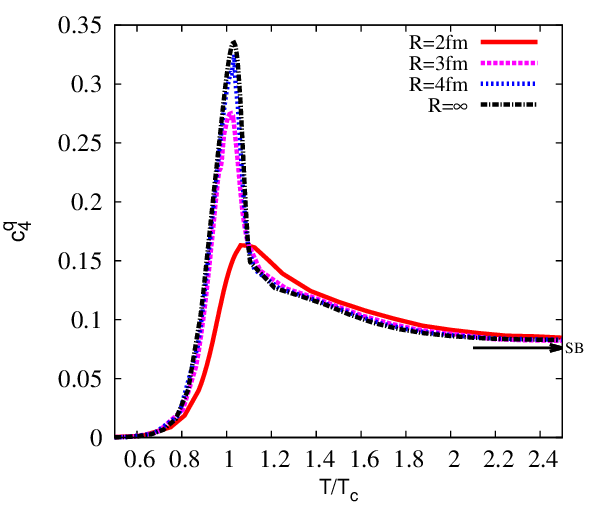}
\includegraphics[scale=0.45]{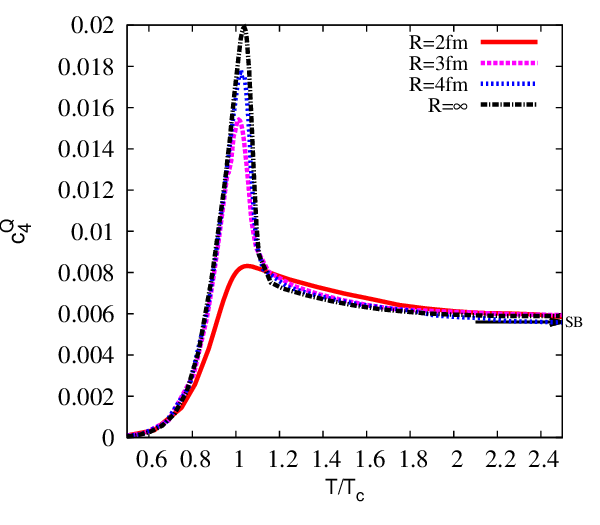}
\includegraphics[scale=0.45]{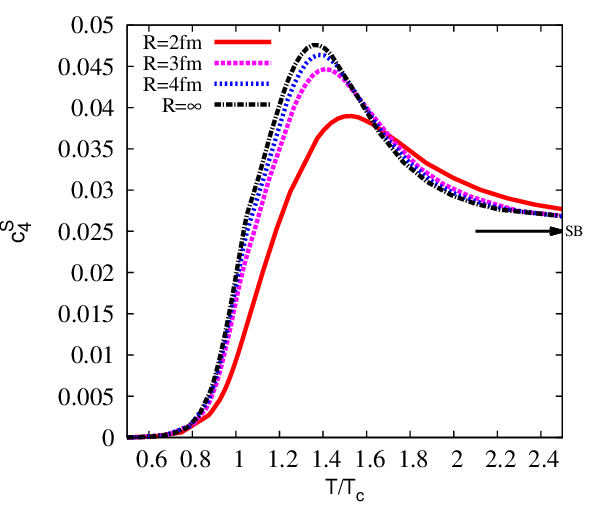}
\end{center}
 \caption{(color online) Variation of second and fourth order
susceptibilities as a function of $T/T_c$ for quark (left column),
electric charge (middle column) and strangeness (right column).  The volume dependence 
of $c_2^q$ has been plotted in the inset.} 
\label{fluct}
\end{figure*}
%%%%%%%%%%%%%%%%%%%%%%%%%%%%%%%%%%%%%%%%%%%%%%%%%%%%%%%%%%%%%%%%%%%%%%%%

The globally conserved charges in the 2+1 flavor matter are the quark
number $q$, the electric charge $Q$ and the strangeness $S$.
The fluctuations of these charges are related to the respective
susceptibilities via the fluctuation-dissipation theorem, and are
obtained as the moments of different orders from the corresponding
chemical potential dependence of the free energy. Close to zero
chemical potential these are given by,
%%%%%%%%%%%%%%%%%%%%%%%%%%%%%%%%%%%%%%%%%%%%%%%%%%%%%%%%%%%%%%%%%%%%%%%%
\begin{equation}
 c^X_n(T)=\frac{1}{n!}\frac{\partial^n(-\Omega(T,\mu_X)/T^4)}
{\partial(\frac{\mu_X}{T})^n}|_{\mu_X=0}
\end{equation}
%%%%%%%%%%%%%%%%%%%%%%%%%%%%%%%%%%%%%%%%%%%%%%%%%%%%%%%%%%%%%%%%%%%%%%%%
where, $X$ stands for either of $q$, $Q$ or $S$. These susceptibilities
have been computed in first principle QCD calculations on the
lattice~\cite{Gottieb_prl,Gottieb_prd,Alton,Gavai_prd1,
HOTQCD_08,Cheng_09,WB_12,HOTQCD_12} as well as in calculations with hard
thermal loops~\cite{Blaizot,purnendu1,jiang,HTLPT_Najmul,
najmul,HMS1_2013,HAMSS_2014,HBAMSS_2014,Ichihara} for large system
sizes. Suitable estimates of these fluctuations are also given by
various QCD inspired models (see e.g.~\cite{Ghosh1,SMukherjee1,
Ghosh2,Ghosh4,sasaki_redlich,FLW_2010,roessner1,Bhat5,
roessner2,schaefer1,schaefer2,schaefer3,schaefer4}). In the context of
finite volumes a study of fluctuations was done by some of us in
Ref.~\cite{Bhattacharyya2}. Here we present the corresponding results in
a 2+1 flavor PNJL model.

To evaluate the fluctuations in the PNJL model, we first compute the
thermodynamic potential $\Omega$ at any particular temperature and with
chemical potentials in a range of -300 MeV$\leq~\mu_X ~\leq$300 MeV with
an interval of 0.1MeV. The scaled thermodynamic potential
$\frac{\Omega}{T^4}$, is then expanded in a Taylor series around
$\mu_X/T$=0. The fluctuations of various orders may be extracted
directly from the coefficients of the series expansion. We have in fact
fitted the scaled thermodynamic potential with a truncated Taylor series
using the GNUPLOT software~\cite{Ghosh1}. To obtain the best fit
parameters up to the $4^{th}$ order, we have checked the least squares
by varying the order of polynomials from 6 to 10 as well as varying the
range of $\mu_X/T$ around zero. The final least squares for the lowest
system size were less than $10^{-9}$ and much smaller ($\sim 10^{-11}$)
for R=3fm, 4fm and $\infty$. The whole procedure was repeated for the
different temperatures in the range of $0.5 < \frac{T}{T_c} < 2.5$.

In fig.~\ref{fluct} we present the $2^{nd}$ and $4^{th}$ order
susceptibilities for all the conserved charges as a function of
$\frac{T}{T_c}$.  From the inset we observe the saturation of $c_2^q$
with increasing system size around $T_c$. As in the case of 2 flavor
system~\cite{Bhattacharyya2}, the qualitative behavior of the
susceptibilities for small system size is similar to that for infinite
volumes, though quantitatively there are significant finite size
effects.  These effects are clearly visible for the $2^{nd}$ order
susceptibilities in the range of $0.8T_c < T < 2 T_c$, and for the
$4^{th}$ order susceptibilities in the range of $0.8T_c < T < 1.2 Tc$.
Therefore if the experimentally produced fireballs thermalize and cool
down to a temperature close to $T_c$, one should consider freeze-out
volume as a fitting parameter along with the temperature and various
chemical potentials to fit the freeze-out multiplicity ratios (see
e.g.~\cite{munzinger} and references therein).  Within the ambit of the
present model, we find that the finite volume effects would be important
if the system freezes out with a size less than $R \sim 5fm$.

%%%%%%%%%%%%%%%%%%%%%%%%%%%%%%%%%%%%%%%%%%%%%%%%%%%%%%%%%%%%%%%%%%%%%%%%
\begin{figure}[!htb] \begin{center}
\includegraphics[scale=0.42]{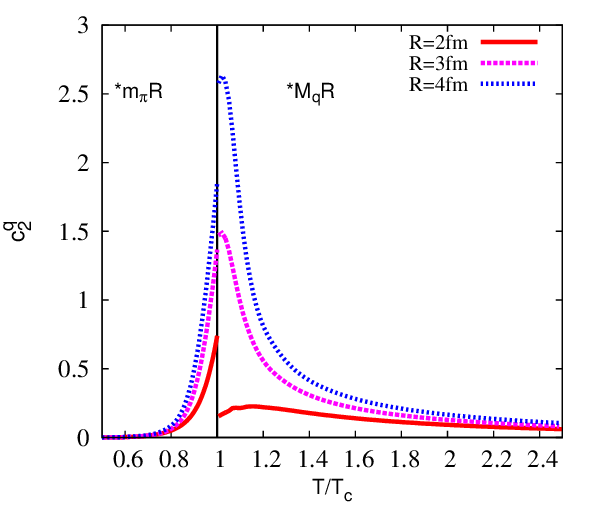}
\includegraphics[scale=0.42]{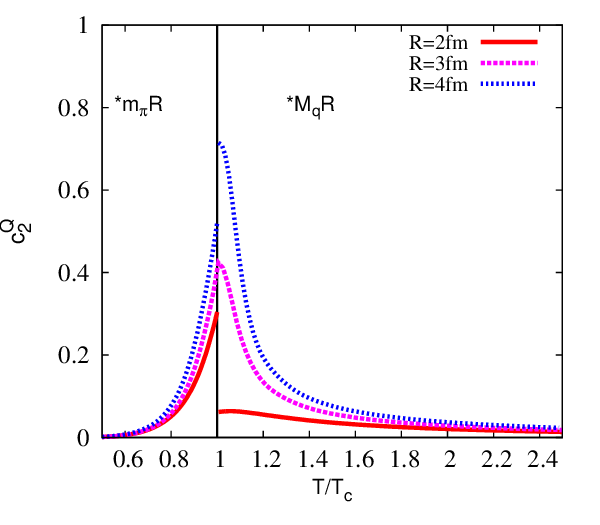}
\includegraphics[scale=0.42]{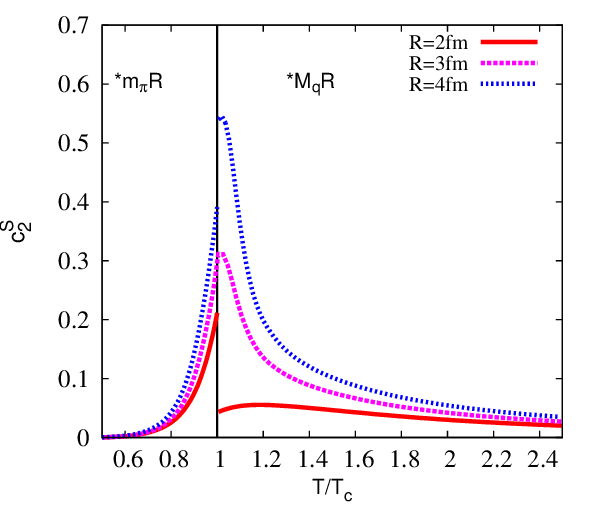} \end{center}
\caption{(color online) Scaling of susceptibilities with $m_{\pi}$ below
$T_c$ and with $M_q$ above $T_c$ as a function of $T/T_c$.  } \label{masssc}
\end{figure}
%%%%%%%%%%%%%%%%%%%%%%%%%%%%%%%%%%%%%%%%%%%%%%%%%%%%%%%%%%%%%%%%%%%%%%%%

To find the relevant length scales in the different temperature
regimes we have also checked the scaling of the susceptibilities with
the corresponding mass scales in the model. For $T<T_c$ the relevant
length scale is the temperature dependent pion mass ($m_{\pi}R$) and
for $T>T_c$ the corresponding scale is the temperature dependent
constituent quark mass ($M_qR$, where $M_q$ is the average constituent
mass for the three flavors). For demonstration we have shown these
scaling behavior in fig.~\ref{masssc}. The relevant masses
were reported by some of us in~\cite{Bhattacharyya1}. We find that for
very low temperatures as well as very high temperatures the mass scaling
is satisfied. But in a significant temperature range close to $T_c$ the
scaling is violated.

%%%%%%%%%%%%%%%%%%%%%%%%%%%%%%%%%%%%%%%%%%%%%%%%%%%%%%%%%%%%%%%%%%%%%%%%
\begin{figure}[!h]
\begin{center}
\includegraphics[scale=0.42]{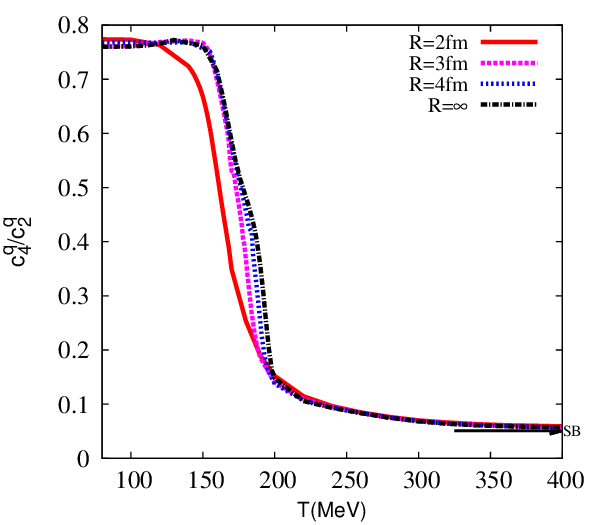}
\includegraphics[scale=0.42]{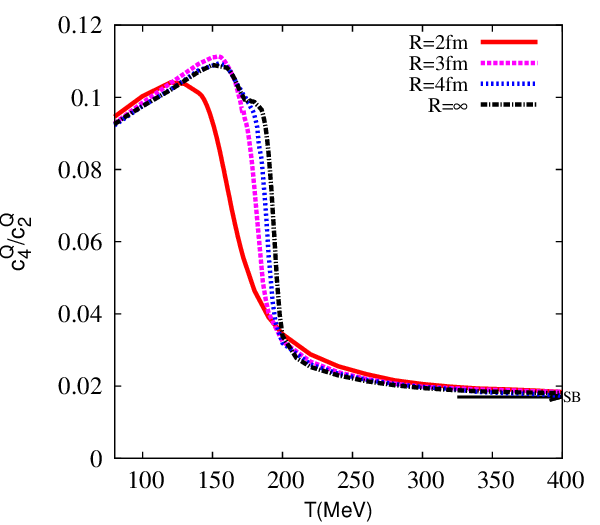}
\includegraphics[scale=0.42]{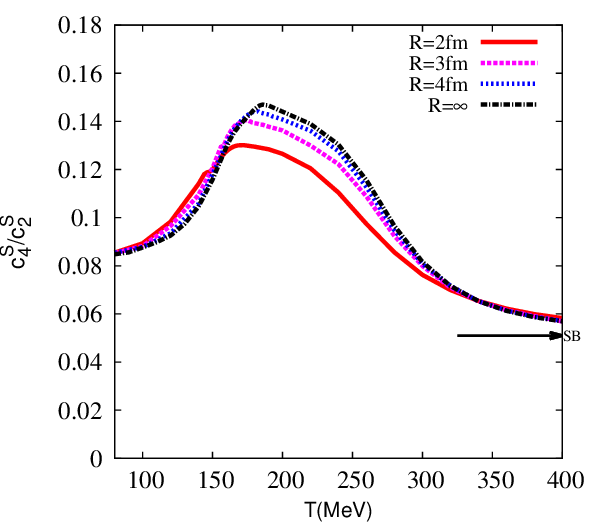}
\end{center}
 \caption{(color online) Ratio of fourth order to second order
susceptibilities for quark, electric charge and strangeness chemical
potentials as a function of $T/T_c$. } 
\label{R42}
\end{figure}

%%%%%%%%%%%%%%%%%%%%%%%%%%%%%%%%%%%%%%%%%%%%%%%%%%%%%%%%%%%%%%%%%%%%%%%%

It should be noted here that for comparing theoretical computations
like in the thermal model~\cite{munzinger}, lattice QCD~\cite{allton},
etc. with experimental data, one usually considers ratios of thermodynamic
observables. This helps in situations where the volume is not directly
measured and the free energy is assumed to be simply proportional to the
volume of the system and therefore scales out in the ratios.
In fact it has been argued~\cite{allton} that in such a situation
the susceptibility ratios are expected to be not only
spectrum independent but dependent only on the ratio of $\mu_q/T$.
Therefore Lattice QCD results corroborates hadron resonance gas results
very well for temperatures below $T_c$. With this expectation, attempts
to match experimentally obtained fluctuation ratios with theoretical
predictions are also undertaken~\cite{aggarwal}. On the other hand some
of us have reported that a finite size system of hadron gas may show
some departure from volume scaling essentially in the electric charge
sector~\cite{Subhasis}. Therefore it is important to check the extent of
volume scaling in the PNJL model. In this regard what we find is that
the susceptibilities themselves
being derivatives of the free energy density are still volume dependent.
Therefore the free energy is not necessarily proportional to the volume
of the system. So a correlated measurement of the particle 
multiplicities and the various susceptibilities may provide suitable
estimates of the finite sizes of the produced strongly interacting
matter with $R < 5 fm$. Recently, with the assumption of volume scaling,
an estimate of system size has been extracted from the fluctuation
and correlations measured by the ALICE collaboration at $\sqrt{s}=2.76$
TeV confronting with the corresponding computations in Lattice QCD
~\cite{munzinger2}.

%%%%%%%%%%%%%%%%%%%%%%%%%%%%%%%%%%%%%%%%%%%%%%%%%%%%%%%%%%%%%%%%%%%%%%%%
\begin{figure}[t]
\begin{center}
 \includegraphics[scale=0.42]{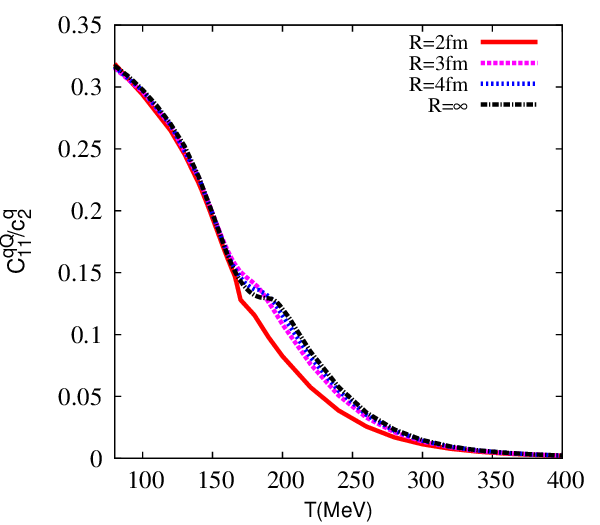}
 \includegraphics[scale=0.42]{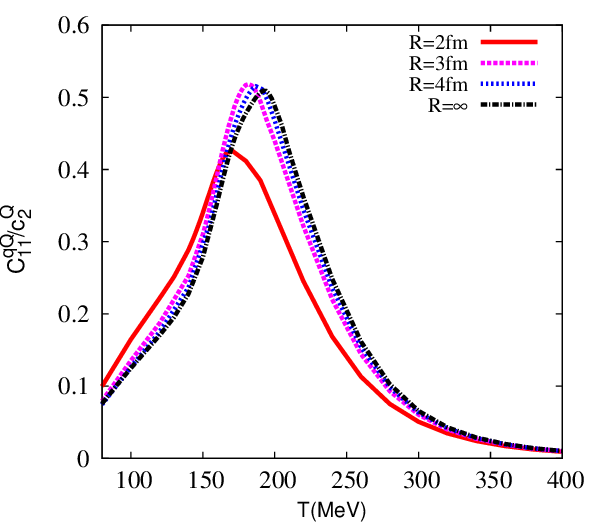}
 \includegraphics[scale=0.42]{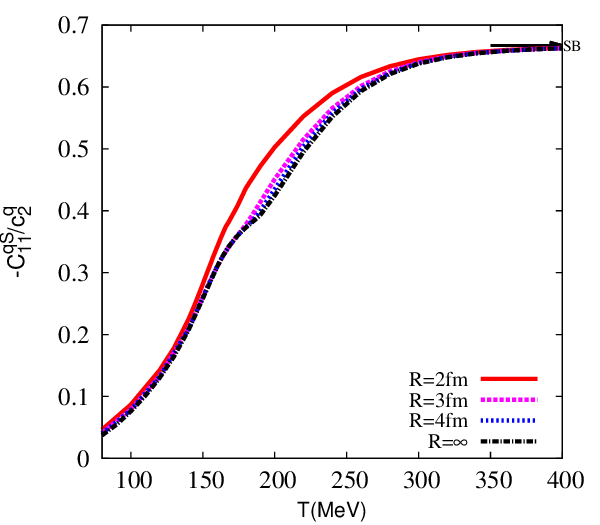}
\end{center}
 \caption{(color online) Ratios of various correlations with fluctuations
as a function of temperature. } 
\label{ratiocorrel}
\end{figure}
%%%%%%%%%%%%%%%%%%%%%%%%%%%%%%%%%%%%%%%%%%%%%%%%%%%%%%%%%%%%%%%%%%%%%%%%

%%%%%%%%%%%%%%%%%%%%%%%%%%%%%%%%%%%%%%%%%%%%%%%%%%%%%%%%%%%%%%%%%%%%%%%%
\begin{figure*}[!htb]
\begin{center}
 \includegraphics[scale=0.45]{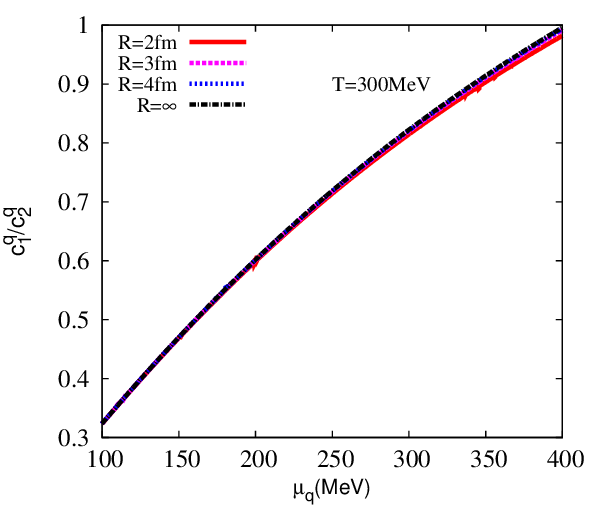}
 \includegraphics[scale=0.45]{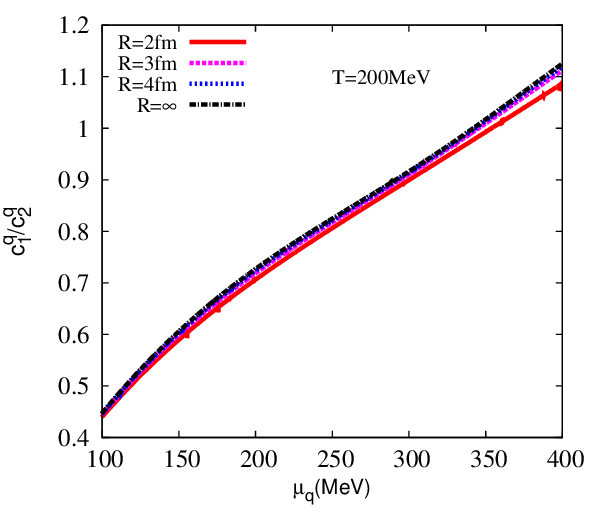}
 \includegraphics[scale=0.45]{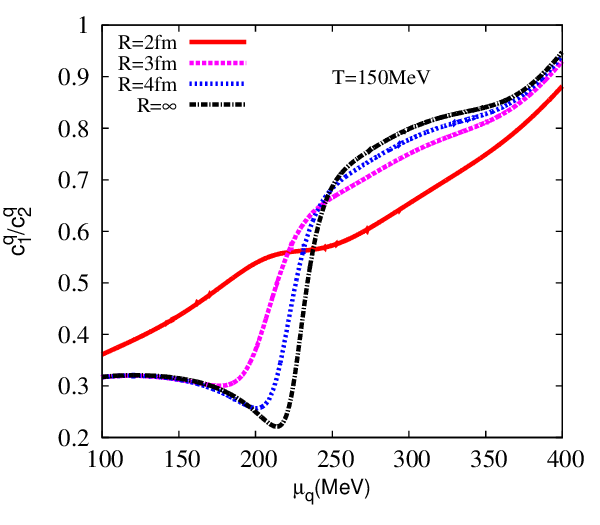}
 \includegraphics[scale=0.45]{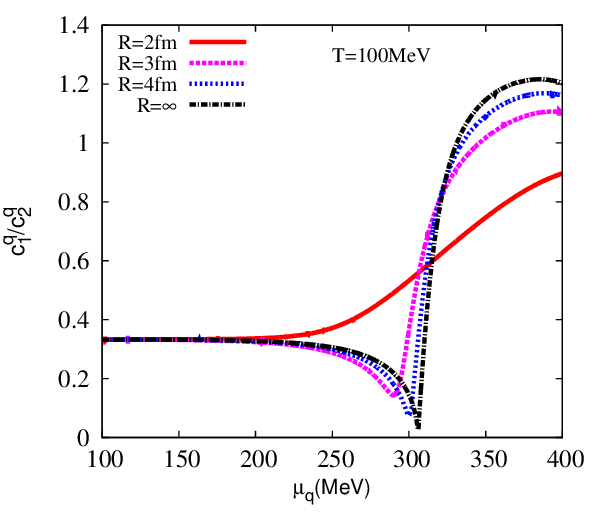}
 \includegraphics[scale=0.45]{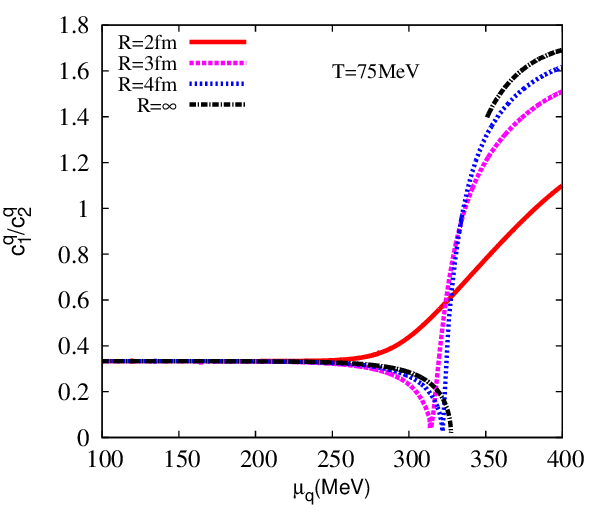}
 \includegraphics[scale=0.45]{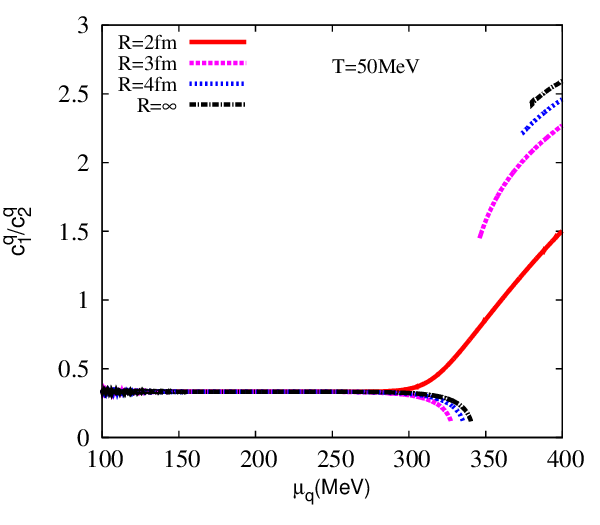}
\end{center}
 \caption{(color online) Ratios of net quark number density to the
quark number fluctuations as functions of quark chemical potential
at different temperatures.} 
\label{ratio_finitemu}
\end{figure*}
%%%%%%%%%%%%%%%%%%%%%%%%%%%%%%%%%%%%%%%%%%%%%%%%%%%%%%%%%%%%%%%%%%%%%%%%

On the other hand if the multiplicity ratios and ratios of fluctuations
are to be employed to make contact between theory and experiment then
it is important to find out if the volume scaling violation occurs for
such ratios. Otherwise no unique value of the system volume may be found.
In fig.~(\ref{R42}) we have plotted the ratios of fourth to second order
fluctuations for different conserved charges as a function of
temperature. It may be observed that the scaling of the ratios with the
system volume holds good away from the crossover region. For all the
conserved charges there is a significant violation of the volume scaling
in a finite window of temperature. For the quark and
electric charge sector the magnitude of violation is high compared to
the strangeness sector, which however shows the scaling violation for
the largest temperature range.

Let us now carry out a study on ratios of leading order correlations 
to fluctuations among different conserved charges. 
The correlations are obtained as,
%%%%%%%%%%%%%%%%%%%%%%%%%%%%%%%%%%%%%%%%%%%%%%%%%%%%%%%%%%%%%%%%%%%%%%%%
\begin{equation}
 C^{X,Y}_{i,j}=\frac{1}{i!j!}\frac{\partial^{i+j}(-\Omega/T^4)}
{(\partial(\frac{\mu_X}{T})^i)(\partial(\frac{\mu_Y}{T})^j)}
\end{equation}
%%%%%%%%%%%%%%%%%%%%%%%%%%%%%%%%%%%%%%%%%%%%%%%%%%%%%%%%%%%%%%%%%%%%%%%%
where, $X$ and $Y$ stands for $q$, $Q$ and $S$ with $X\neq Y$. We
compute the correlation coefficients using numerical differentiation. 
We obtain the thermodynamic potential $\Omega$ at any given temperature
within the range -50MeV$\leq \mu_X ,\mu_Y \leq$50MeV in an interval
of 0.1 MeV. For obtaining the derivative 
$\frac{\partial^2(-\Omega/T^4)}{\partial(\mu_X/T)\partial(\mu_Y/T)}$, 
we compute it for various values of $\Delta(\mu_X/T)=\Delta(\mu_Y/T)$
and consider the value when the derivative reaches a saturation.
We repeat the whole procedure for the entire temperature range so as to
obtain first order correlations of conserved charges as functions of
temperature around zero chemical potentials. 

In fig.~\ref{ratiocorrel} we have plotted the ratios of correlations 
to fluctuations for different conserved charges. The volume scaling 
violation in this case is quite small for all the sectors except for 
the smallest system size. We note that here the correlations 
and fluctuations are derivatives of same order of the free energy. 

The fluctuations of the conserved charges can also be extracted for
non-zero chemical potentials. Here we have considered net quark density
$c_1^q$ and quark number fluctuations $c_2^q$ for simplicity. These
quantities are also the simplest to analyze from the experimental data
at least in terms of net proton number and its
fluctuations~\cite{aggarwal}. We compute the thermodynamic potential as
a function of $\mu_q$ in a range of $0<\mu_q<400$MeV with an interval of
$0.1$MeV for a set of temperatures. Then using the method of numerical
derivative we obtain the observables.

The quark number density itself was found to have a volume dependence.
The ratio $c_1^q/c_2^q$ is displayed in fig.~\ref{ratio_finitemu} for a
few representative values of temperature.  Since the quark number is
very small for low values of $\mu_q$ we present our results in the range
100MeV$<\mu_q<$400MeV.  For high temperatures the ratio is monotonically
increasing. On the other hand for $T< 200$ MeV we observe a dip in a
certain window of $\mu_q$. Incidentally these windows of $\mu_q$
correspond to those close to the phase boundary~\cite{Bhattacharyya1}.
The dips occur due to the shooting up of the quark number fluctuations
in these regions. For $T < 100$ MeV there is even a discontinuity
indicating the existence of the first order line in the phase diagram
for the corresponding system size.  We observe that for high
temperatures the volume scaling is maintained even for large chemical
potentials. However as we go down below $T\sim200$ MeV the volume
scaling is violated in the transition region of the corresponding
temperature. Thus we may infer that the violation of volume scaling if
any, always occur close to the transition region, where large
correlation lengths come into play and lead to the separate finite size
behavior of the derivatives of the free energy.

\section{Conclusion}

To conclude, we have studied the volume dependence of the free energy
density of 2+1 flavor strongly interacting matter in terms of various
second and fourth order fluctuations of the conserved charges, as well
as the quark number density at finite temperature and chemical
potentials. We observed volume scaling violations at two levels. Firstly
the free energy density is itself volume dependent for $R < 5 fm$ within
a certain range of $T$ and $\mu_q$, as given by the behavior of the
susceptibilities. Secondly the ratio of these derivatives themselves
show violation of volume scaling in a small window of $T$ and $\mu_q$
all along the transition region in the $T-\mu_q$ phase diagram.
The addition of one more flavor in our treatment did not change any
qualitative features of the quark number fluctuations compared to the
the 2 flavor system~\cite{Bhattacharyya2}, though the quantitative
results are significantly different. The 2nd order strangeness
susceptibility has a substantial contribution to the quark number
fluctuation. On the other hand the 4th order strangeness susceptibility
has a sub-dominant contibution to the corresponding quark number
susceptibility. As a result the volume scaling violation in the ratio of
4th order to 2nd order quark number fluctuations is somewhat smaller in
2+1 flavor system compared to the 2 flavor system. However a cleaner
signal of volume scaling violation is now observed in the ratios of
strangeness susceptibilities themselves, which is sustained in a much
wider range of temperatures. We have also been able to extract various
charge correlations. The ratios of these correlations with the 2nd order
fluctuations violate volume scaling only for very small system sizes.
We have further extended our studies to non-zero chemical potentials.
An interesting feature observed is that the first order phase transition
nature is itself manifestly dependent on the system size, as can be seen
from the quark number to quark susceptibility ratios. For very small
systems the first order phase boundary is completely washed out.
Further investigation in these directions are required with the
computation of higher order moments that we wish to carry out in future.

\acknowledgments
Authors thank Council for Scientific and Industrial Research (CSIR),
Department of Science and Technology (DST), University Grants 
Commission (UGC) and Alexander von Humboldt
(AvH) Foundation for support.

\end{document}